\documentclass{article}
\usepackage{spconf,amsmath,graphicx,hyperref,booktabs}
\usepackage{enumitem}
\usepackage{subcaption}
\usepackage[english]{babel}
\usepackage{microtype}
\usepackage{graphicx}
\usepackage{bm}              
\usepackage{color}
\usepackage{array} 
\newcolumntype{C}{>{\centering\arraybackslash}X}
\newcolumntype{Y}{>{\centering\arraybackslash}X}
\usepackage{tabularx}
\usepackage{booktabs}
\usepackage[table]{xcolor}
\usepackage{siunitx}
\usepackage{makecell}
\usepackage{multirow}
\usepackage{array}     
\usepackage{cite}
\usepackage{makecell}
\title{Adaptive Per-Channel Energy Normalization Front-end for Robust Audio Signal Processing}
\name{Hanyu Meng$^{\star}$, Vidhyasaharan Sethu$^{\star}$, Eliathamby Ambikairajah$^{\star}$, Qiquan Zhang$^{\dagger}$, Haizhou Li$^{\ddagger}$\thanks{\textcolor{black}{This research work is funded by the Australian Research Council (ARC) Discovery Grant DP210101228. The code is available at}: \url{https://github.com/Hanyu-Meng/LEAF-APCEN}}}

\address{
  $^{\star}$The University of New South Wales, Sydney, Australia \\
  $^{\dagger}$Tongyi Speech Lab, Alibaba Group, China \\
  ${\ddagger}$School of Artificial Intelligence, The Chinese University of Hong Kong, Shenzhen, China\\
}
\begin{document}
\ninept
%
\maketitle
\vspace{-2mm}
\begin{abstract}
In audio signal processing, learnable front-ends have shown strong performance across diverse tasks by optimizing task-specific representation. However, their parameters remain fixed once trained, lacking flexibility during inference and limiting robustness under dynamic complex acoustic environments. In this paper, we introduce a novel adaptive paradigm for audio front-ends that replaces static parameterization with a closed-loop neural controller. Specifically, we simplify the learnable front-end LEAF architecture and integrate a neural controller for adaptive representation via dynamically tuning Per-Channel Energy Normalization. The neural controller leverages both the current and the buffered past subband energies to enable input-dependent adaptation during inference. Experimental results on multiple audio classification tasks demonstrate that the proposed adaptive front-end consistently outperforms prior fixed and learnable front-ends under both clean and complex acoustic conditions. These results highlight neural adaptability as a promising direction for the next generation of audio front-ends.
\end{abstract}

\vspace{-4mm}
\section{Introduction}
\begin{figure}[!ht]
  \centering
  \begin{subfigure}[b]{\linewidth}
    \centering
    \includegraphics[width=0.95\linewidth]{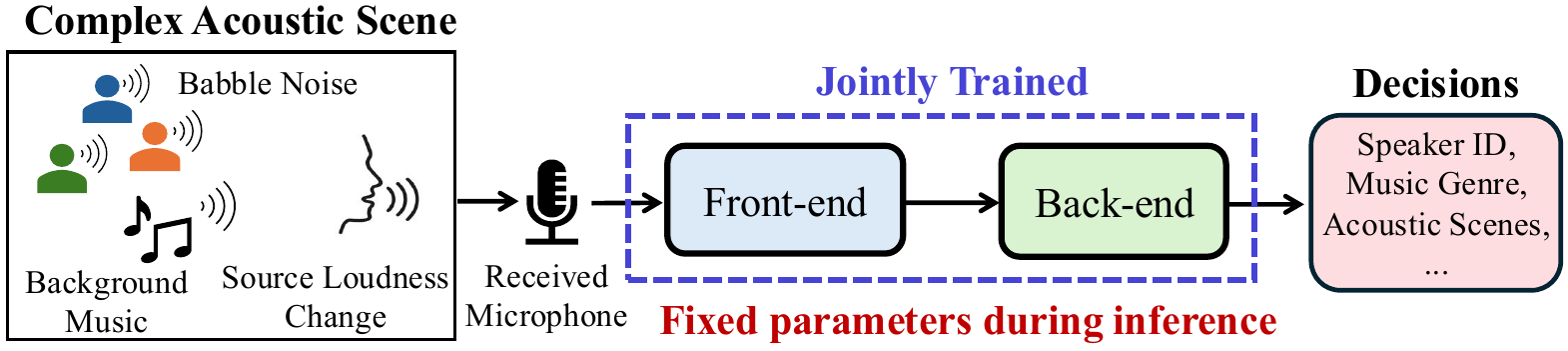}
    \subcaption{Learnable Front-end Paradigm}
    \label{fig:intro_A}
  \end{subfigure}
  \begin{subfigure}[b]{\linewidth}
    \centering
    \includegraphics[width=0.95\linewidth]{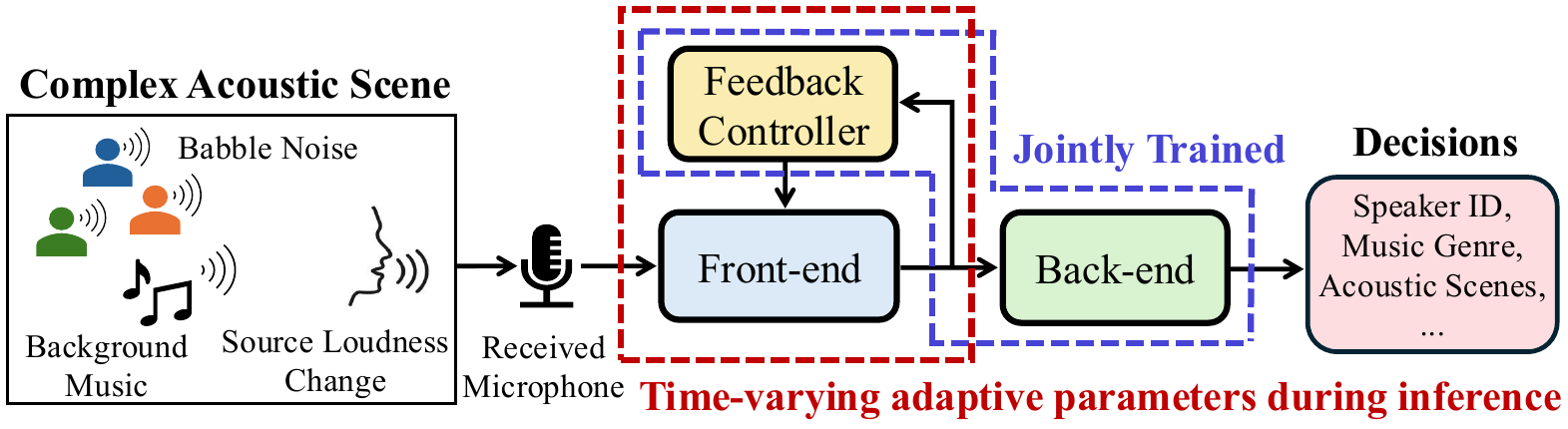}
    \subcaption{Adaptive Front-end Paradigm}
    \label{fig:intro_B}
  \end{subfigure}
  \caption{Comparison between two classes of front-ends: (a) learnable and (b) adaptive. The proposed front-end adopts an adaptive filterbank with dynamic range compression, producing a spectrogram-like representation for downstream classification.}
  \vspace{-5.5mm}
  \label{fig:intro}
\end{figure}

\vspace{-1mm}
\label{sec:intro}
Effective audio front-ends are crucial for extracting informative representations that directly impact the performance of downstream audio processing tasks. Over the past decade, with the rapid development of deep learning, the front-end design paradigm has shifted from fixed handcrafted features, such as the Mel filterbank~\cite{davis1980mfcc} or constant Q-transform~\cite{cqt}, to learnable front-ends employing convolutional filters~\cite{learnable_cnn,learnable_cnn2,11184233}, or parametric neural filters~\cite{LEAF,efficient_leaf,sincnet,harmonic_filter,liu2024learning}. These learnable front-ends are trained jointly with the back-end network, enabling task-specific representations to be optimized for the downstream objective as shown in Fig.~\ref{fig:intro_A}. Subsequent research has focused on improving their efficiency, stability, and applicability across tasks~\cite{efficient_leaf,Leaf_do_not_learn,meng23_interspeech,fastaudio_learnable_FE_spoof_detect}, thereby overcoming the inflexibility of fixed feature extractors.  

Despite their success, a fundamental limitation remains: the parameters of learnable front-ends are fixed once training is complete, making them unable to adapt to varying input conditions (e.g. changing signal content or complex acoustic environments) during inference~\cite{Leaf_do_not_learn}. To address this, recent work has explored adaptive front-ends that incorporate feedback mechanisms to change the filterbank's Q factor in the spectral decomposition process for simulating the human auditory system~\cite{buddhi_AdaFE,AdaFE_Qiquan,similar_AdaFE}. As illustrated in Fig.~\ref{fig:intro_B}, these adaptive front-ends train a feedback controller alongside the back-end, enabling time-varying parameterization of the front-end during inference. This allows the system to maintain robust performance in dynamic acoustic conditions such as babble noise, background music, or time-varying loudness.

Building on the insight that audio front-ends should be adaptive~\cite{AdaFE_Qiquan}, and motivated by prior investigations of the adaptability potential of Per-Channel Energy Normalization (PCEN) in the LEAF framework~\cite{meng23_interspeech}, we propose a novel adaptive LEAF, termed \textbf{LEAF-APCEN}. Unlike previous biologically inspired adaptive frameworks~\cite{AdaFE_Qiquan,buddhi_AdaFE}, which adaptively reshape spectral decomposition filters (change Q-factors of Gabor filters), our approach focuses on the PCEN stage, dynamically adjusting two key parameters for time-varying dynamic range compression. To the best of our knowledge, this is the first adaptive front-end to realise subband energy normalization through a neural controller. The main contributions of this work are as follows:
\begin{figure*}[!ht]
  \centering
  \begin{subfigure}[b]{0.66\textwidth}
    \centering
    \includegraphics[width=\linewidth]{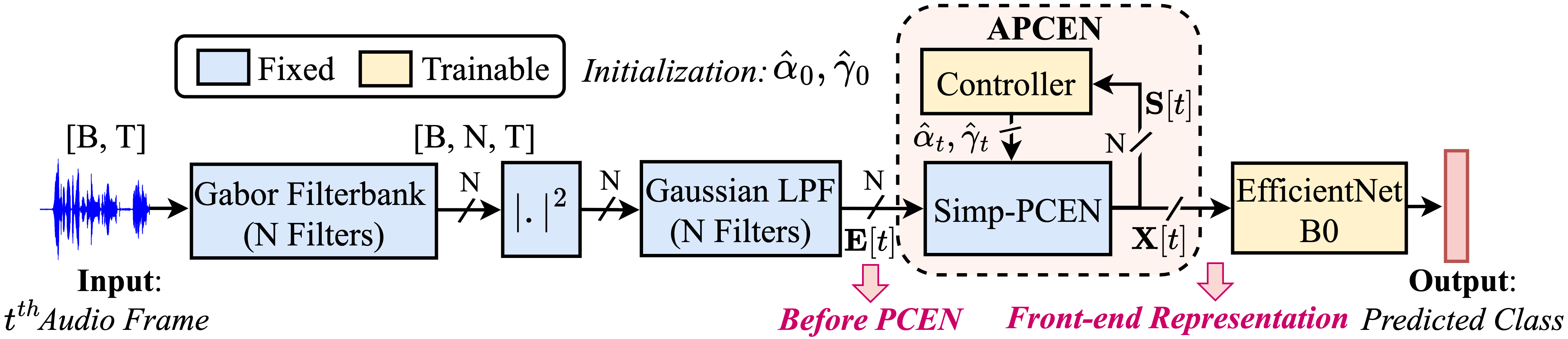}
    \caption{The Structure of LEAF-APCEN and back-end classifier.}
    \label{fig:overview}
  \end{subfigure}
  \hfill
  \begin{subfigure}[b]{0.33\textwidth}
    \centering
    \includegraphics[width=\linewidth]{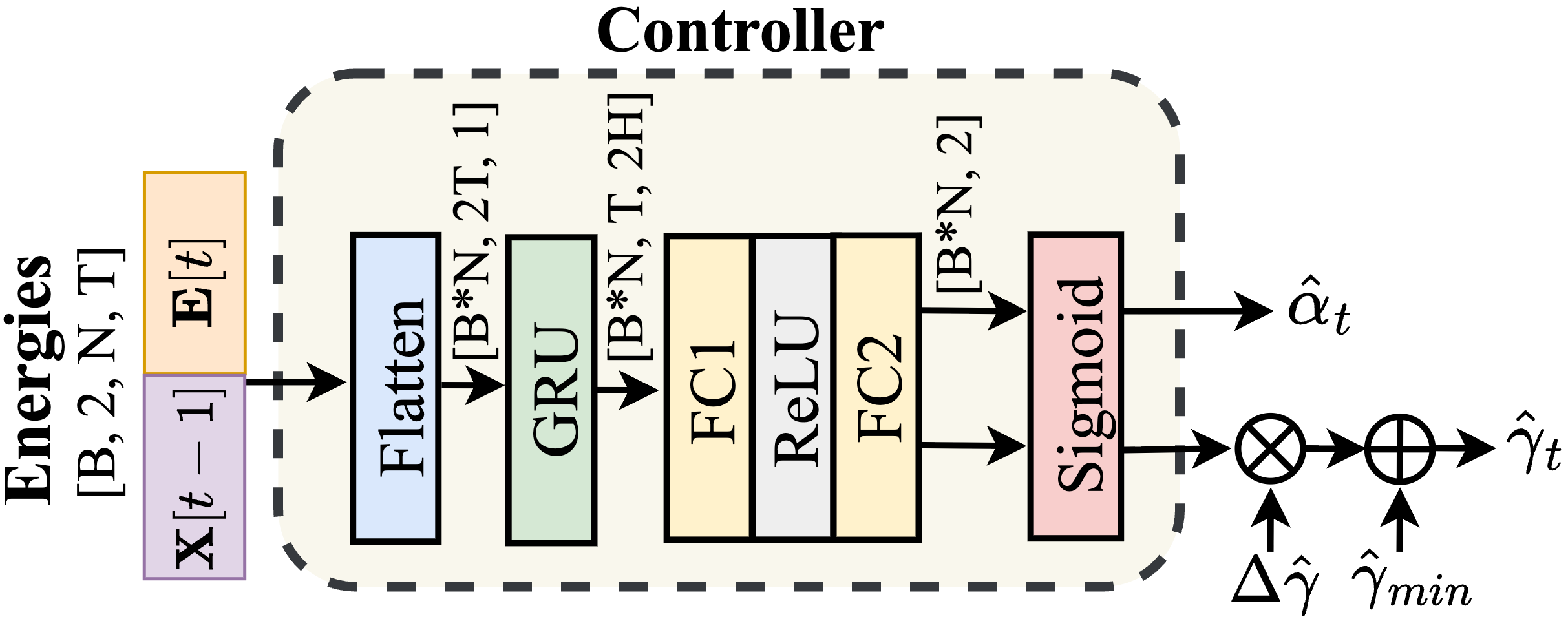}
    \caption{Neural Adaptive Controller.}
    \label{fig:controller}
  \end{subfigure}
  \caption{Proposed Model Overview. ($B$: Batch size, $N$: Filter numbers, $T$: Frame length, $H$: Hidden states).}
  \label{fig:overview_controller}
\end{figure*}
\begin{itemize}[itemsep=0pt, topsep=2pt, leftmargin=*]
    \item \textbf{Simplified PCEN.} We reformulate the original PCEN~\cite{original_pcen} by reducing its four learnable parameters to two (SimpPCEN), achieving a more efficient design without degrading performance.
    \item \textbf{Adaptive PCEN.} We introduce LEAF-APCEN, an adaptive PCEN front-end built on SimpPCEN and the LEAF framework, where a neural controller dynamically adjusts parameters during inference to enable time–frequency dependent dynamic range compression for robust representation.
    \item \textbf{Neural Adaptive Controller.} We design a lightweight neural controller that takes current and previously processed subband energies as input to effectively regulate SimpPCEN parameters.
\end{itemize}
\section{Proposed Method}
\label{sec:method}
\subsection{LEAF-APCEN}
\label{sec:2.1}
The overview of our proposed \textbf{A}daptive \textbf{P}er-\textbf{C}hannel \textbf{E}nergy \textbf{N}ormalisation (APCEN) mechanism is illustrated in Fig.~\ref{fig:overview}. Our model, referred to as LEAF-APCEN, is built on top of the LEAF framework, a widely used learnable audio front-end~\cite{LEAF}. LEAF employs a fixed Gabor filterbank with $N$ filters placed on a Mel-scale for auditory perception-aware spectral decomposition of the input audio. This is followed by subband Gaussian Low-Pass Filters (LPFs) that smooth temporal fluctuations. At the last stage of the LEAF-APCEN, unlike the original LEAF, which uses a static PCEN component, our approach introduces a novel APCEN module. This module performs adaptive dynamic range compression by adjusting gains based on the input signal's energy. This adaptation is achieved via a dedicated neural feedback controller, which is detailed in Section~\ref{sec:2.3}.

Following recent studies~\cite{meng23_interspeech,Leaf_do_not_learn} that have shown the first two components of LEAF (the filterbank and smoothing layers) do not need to be learnable for strong performance, we keep them fixed. This allows us to focus our adaptation efforts exclusively on the PCEN stage. Furthermore, we simplify the original PCEN algorithm, which typically requires four learnable parameters, into a simplified version (Simp-PCEN) with only two tunable parameters, as described in Section~\ref{sec:2.2}. 

For the adaptation, we draw inspiration from prior work~\cite{buddhi_AdaFE,AdaFE_Qiquan} on adaptive front-ends by introducing a trainable neural adaptive controller. This controller dynamically tunes the two parameters of Simp-PCEN, effectively modeling the frequency-dependent and time-varying gain control found in the human auditory system. It is crucial to note that in our proposed APCEN module, the Simp-PCEN parameters are not directly trainable via backpropagation; their values are adjusted only by the neural adaptive controller.
\subsection{Simplified Per-Channel Energy Normalization (SimpPCEN)}
\label{sec:2.2}
As mentioned in the previous section, we simplified the original PCEN algorithm into a version we call Simp-PCEN, which uses only two parameters. The original PCEN for the $n^{th}$ frame and $i^{th}$ filter can be written as~\cite{original_pcen}:
\begin{equation}
\mathrm{PCEN}[n,i] = \left( \frac{E[n,i]}{(M[n,i] + \epsilon)^{\bm{\alpha}_i}} + \bm{\delta}_i \right)^{\bm{\gamma}_i} - \bm{\delta}_i^{\bm{\gamma}_i},
\label{eq:pcen}
\end{equation}
where $M[n,i]$ is the smoothed energy estimate for that subband, calculated as:
\begin{equation}
M[n,i] = \bm{s}_i E[n,i] + (1 - \bm{s}_i) M[n-1,i].
\end{equation}

The bolded parameters, $(\bm{s}_i, \bm{\alpha}_i, \bm{\delta}_i, \bm{\gamma}_i)$, are learnable for each frequency band. Following the original PCEN study, we set their initial values to $s_0=0.04$, $\alpha_0=0.96$, $\delta_0=2$, and $\gamma_0=0.5$. 

Our simplified equation for SimpPCEN is as follows:
\begin{equation}
\mathrm{SimpPCEN}[n,i] = \frac{ \left(E[n,i]\right)^{\hat{\bm{\gamma}}_i} }{ \left(M[n,i] + \epsilon\right)^{\hat{\bm{\alpha}}_i} }.
\label{eq:simp-pcen}
\end{equation}
Compared to the original PCEN, we removed the offset parameter, $\bm{\delta}_i$, and set the smoothing factor, $\bm{s}_i$, to a fixed initial value, $s_0$. To ensure consistency, the simplified parameters are initialized as $\hat{\gamma}_0 = \gamma_0 = 0.5$ and 
$\hat{\alpha}_0 = \alpha_0 \gamma_0 = 0.48$, derived from the original PCEN initial values and the reformed equation.

\begin{figure}[!ht]
  \centering
  \hspace{-0.5cm}
  \includegraphics[width=0.92\columnwidth]{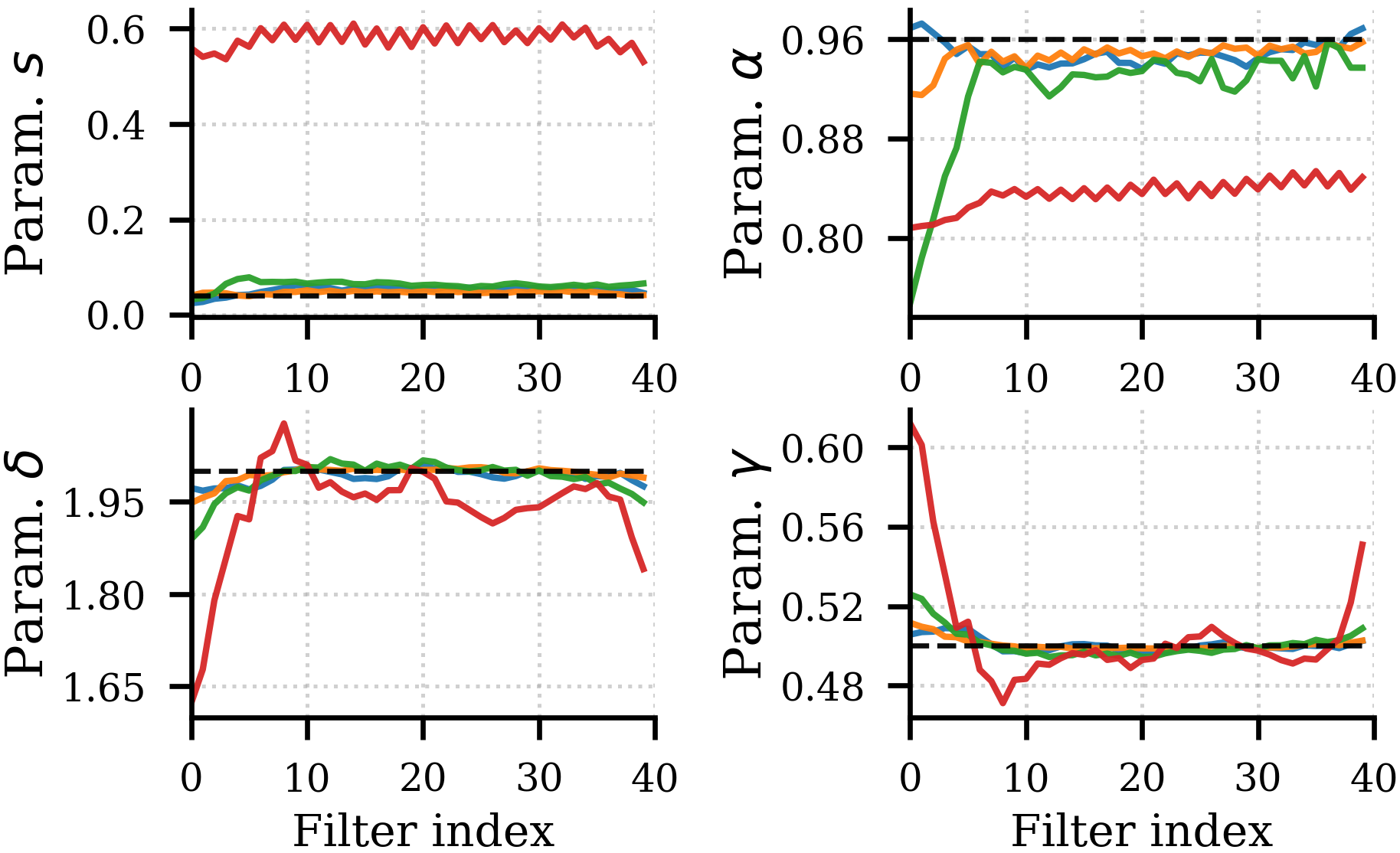}
  \vspace{0.3em} 
  \hspace{0.8cm}
  \includegraphics[width=0.85\columnwidth]{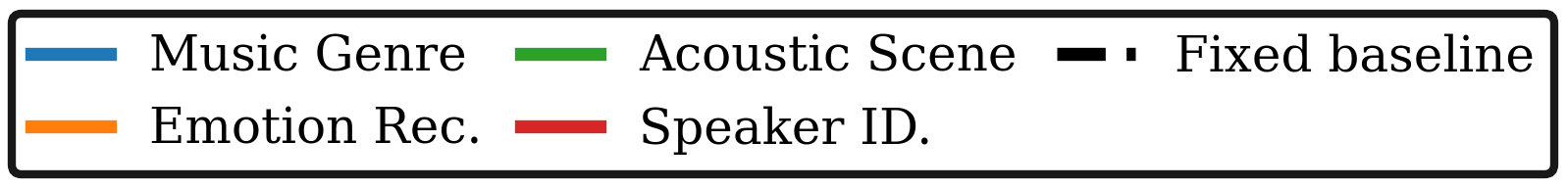}
  \caption{Fully learned PCEN parameter variations across filter indices for different audio classification tasks, compared with their initial values. The four subplots show the learned distributions for $\bm{s}$, $\bm{\alpha}$, $\bm{\delta}$, and $\bm{\gamma}$, respectively.}
  \label{fig:pcen_variation}
\end{figure}

The motivation for this simplification is supported by both empirical observations and task performance analysis. When we compared the learned parameter distributions of the fully learnable PCEN (from the LEAF framework) across the four audio classification tasks, we found that the variations in $\bm{s}$ and $\bm{\delta}$ were relatively minor as shown in Fig.~\ref{fig:pcen_variation}. Even for the speaker identification task, where the parameter distributions differed slightly, the variations in $\bm{s}$ and $\bm{\delta}$ were insufficient to yield a clear performance gain, as shown in Table~\ref{tab:robust_res_clean}. In contrast, the distribution of $\bm{\gamma}_i$ demonstrated that even small changes led to large behavioral shifts. Since the exponent plays a central role in the automatic gain control mechanism~\cite{pcen_explain}, we retained it as a learnable parameter. 

Based on these findings, we simplified PCEN by removing the less influential parameters while retaining the most critical ones. Experimental results in Table~\ref{tab:robust_res_clean} further confirm that LEAF with Simp-PCEN achieves comparable or even better accuracy than LEAF with the original four-parameter PCEN across multiple tasks, demonstrating that our proposed simplification is both effective and necessary.

\subsection{Neural Adaptive Controller}
\label{sec:2.3}
As illustrated in Figure~\ref{fig:controller}, the feedback controller is designed to dynamically adjust the SimpPCEN parameters based on the acoustic context. Its input, $\mathbf{S}[t] = [\mathbf{X}[t-1]; \mathbf{E}[t]]$, is a concatenation of the current frame's input energy vector $\mathbf{E}[t]$ and the previously processed frame's output $\mathbf{X}[t-1]$. This allows the controller to model the temporal dynamics of the subband energies.

The controller's architecture consists of a bidirectional Gated Recurrent Unit (GRU) to capture these time-varying subband energy changes. The output of the GRU is passed to a two-layer Multi-Layer Perceptron (MLP) with ReLU activation, which predicts the two SimpPCEN parameters, $\hat{\alpha}_t$ and $\hat{\gamma}_t$. A sigmoid function is applied to the MLP output to constrain the parameters within the $(0, 1)$ range, consistent with prior work on LEAF and PCEN applications~\cite{efficient_leaf,pcen_application,LEAF, original_pcen}. The neural adaptive controller has a parameter number of $4.32$\,K.

To ensure stability and prevent degenerate cases where $\hat{\gamma}_t$ approaches zero, we apply an affine transformation to its output. We found that the $\hat{\gamma}_t$ parameter is particularly sensitive, so we rescale it using a predefined range $\Delta\hat{\gamma}$ and a minimum value $\hat{\gamma}_{min}$. This transformation helps stabilize the parameter and improve overall model performance. For the detailed numerical settings of the neural structure and hyperparameters, please refer to Section~\ref{sec:3.2}.

\section{Experimental Setups}
\label{sec:experiment}
\subsection{Data Preparation}
\label{sec:3.1}
To thoroughly evaluate our method, we conducted a comprehensive investigation across a broad range of audio signal classification tasks, encompassing environmental sound, music, and speech perception of identity and paralinguistic, consistent with previous studies~\cite{explore_pretrained,AST,AdaFE_Qiquan}. We evaluated our method on four representative audio classification tasks: acoustic scene classification (ESC-50~\cite{esc50}), music genre classification (FMA-Small~\cite{fma}), emotion recognition (CREMA-D~\cite{crema_d}), and speaker identification (VoxCeleb1~\cite{voxceleb}). Dataset statistics are provided in Table~\ref{tab:datasets}.

All models were trained and tested under two conditions: \textbf{(i) clean audio} and \textbf{(ii) the complex acoustic condition datasets} designed to simulate non-stationary environments. The latter was created by adding three types of perturbations and the clean data in equal proportions across training and validation: \textbf{(a) Babble noise}, following~\cite{x_vector_use_musan}, generated by mixing three speakers from the MUSAN dataset~\cite{musan} at random Signal-to-Noise Ratios (SNRs) between 0–15 dB; \textbf{(b) Background music}, sampled from the MUSAN music subset and mixed using the same procedure; and \textbf{(c) Loudness changes}, produced by applying random gains between –8 and +8 dB every 250 ms to keep the Sound Pressure Level (SPL) within –40 to –15 dBFS~\cite{dns_icassp2021}. To ensure fairness, the amount of each condition was balanced across classes.
\begin{table}[!ht]
\centering
\captionsetup{skip=2pt}      
\caption{Statistics of the four datasets used in our experiments, including class count, average duration, and split sizes.}
\label{tab:datasets}
\setlength{\tabcolsep}{3pt} 
\renewcommand{\arraystretch}{0.85} 
\footnotesize                
\begin{tabularx}{\linewidth}{lYYYY}
\toprule
\rowcolor{gray!10}
\textbf{Task} & \textbf{Acous.\ Scene} & \textbf{Music\ Genre} & \textbf{Emo.\ Rec.} & \textbf{Spk.\ ID} \\
\cmidrule(lr){1-5}
\textbf{Dataset} &
\makecell{ESC\mbox{-}50\scriptsize} &
\makecell{FMA\mbox{-}S\scriptsize} &
\makecell{CREMA\mbox{-}D\scriptsize} &
\makecell{VoxCeleb1\scriptsize} \\
\midrule
\# classes          & 50     & 8      & 6      & 1251 \\
Average duration    & 5.0\,s & 30.0\,s & 2.5\,s & 8.2\,s \\
\midrule
\# train clips      & \multirow{3}{*}{\makecell[c]{5 folds\\ 2000}} & 6400  & 5155  & 138316 \\
\# validation clips &                                           & 800   & 732   & 6904   \\
\# test clips       &                                           & 800   & 1551  & 8251   \\
\bottomrule
\end{tabularx}
\end{table}
\vspace{-2mm}
\begin{table*}[!ht]
\centering
\captionsetup{skip=2pt}
\caption{Top-1 accuracy (\%, $\uparrow$) for models trained and evaluated on \textbf{clean} audio ($\uparrow$ indicating higher is better).}
\label{tab:robust_res_clean}
\begin{tabular}{l|c|c|cccc}
\hline
\rowcolor[HTML]{EFEFEF} 
\textbf{Model} & \textbf{Data Type} & \textbf{Front-end Info.} & \textbf{Acous. Scene} & \textbf{Music Genre} & \textbf{Emo. Rec.} & \textbf{Spk. ID.} \\ \hline
Fixed LEAF~\cite{LEAF}               & Clean     & Fixed     & 55.75 & 47.88 & 50.92 & 41.34 \\
LEAF-PCEN~\cite{LEAF,Leaf_do_not_learn} & Clean     & 4 Params. Learnable & 56.75 & 48.38 & 51.58 & 35.49 \\
LEAF-SimpPCEN  & Clean     & 2 Params. Learnable & 57.25 & \textbf{50.63} & 51.97 & 35.37 \\ 
LEAF-APCEN (Proposed) & Clean & 2 Params. Adaptive & \textbf{61.25} & 48.63 & \textbf{59.32} & \textbf{49.84} \\ \hline
\end{tabular}
\end{table*}
\begin{table*}[!ht]
\centering
\captionsetup{skip=2pt}
\caption{Top-1 accuracy (\%, $\uparrow$) with models trained and evaluated on created \textbf{complex acoustic condition} audio.}
\label{tab:robust_res_complex}
\begin{tabular}{l|c|c|cccc}
\hline
\rowcolor[HTML]{EFEFEF} 
\textbf{Model} & \textbf{Data Type} & \textbf{Front-end Info.} & \textbf{Acous. Scene} & \textbf{Music Genre} & \textbf{Emo. Rec.} & \textbf{Spk. ID.} \\ \hline
Fixed LEAF~\cite{LEAF}               & Complex Env.  & Fixed     & 40.00 & 43.13 & 43.97 & 40.69 \\
LEAF-PCEN~\cite{LEAF,Leaf_do_not_learn} & Complex Env.  & 4 Params. Learnable & 38.50 & 45.50 & 45.26 & 34.28 \\
LEAF-SimpPCEN & Complex Env. & 2 Params. Learnable & 39.75 & 43.88 & 44.75 & 34.04 \\ 
LEAF-APCEN (Proposed) & Complex Env. & 2 Params. Adaptive  & \textbf{55.75} & \textbf{46.50} & \textbf{51.97} & \textbf{49.41} \\ \hline
\end{tabular}
\end{table*}
\vspace{-2mm}
\subsection{Model Configurations}
\label{sec:3.2}
We evaluate four LEAF-based front-ends:\textbf{(i) Fixed LEAF}: all front-end parameters are fixed to their initial values. \textbf{(ii) LEAF-PCEN}: the Gabor filterbank and Gaussian LPFs are fixed; only the four PCEN parameters are learnable. \textbf{(iii) LEAF-SimpPCEN}: as above, but replace PCEN with SimpPCEN and learn only its two parameters. \textbf{(iv) LEAF-APCEN}: as above, but use the proposed APCEN in which a learnable neural adaptive feedback controller predicts the two SimpPCEN parameters. These variants span four LEAF front-end paradigms: fixed, four-parameter learnable, two-parameter learnable, and adaptive parameterization.

We follow~\cite{explore_pretrained,AST,AdaFE_Qiquan} to resample all audio to 16\,kHz. Signals are framed with a 25\,ms window and a 10\,ms hop, consistent with LEAF~\cite{LEAF}. The first two stages use $N=40$ Gabor filters and Gaussian LPFs with filter length $L=150$. PCEN and SimpPCEN are initialized as described in Section~\ref{sec:2.2} and the original LEAF~\cite{LEAF}. For APCEN, the adaptive controller is a bidirectional GRU with hidden size $H=32$, followed by two fully connected layers with 32 and 2 output channels, respectively. We constrain $\hat{\gamma_t}$ via $\hat{\gamma}_{\min}=0.2$ and range $\Delta\hat{\gamma}=0.8$. As for the back-end classifier, we chose EfficientNetB0~\cite{efficientnet}.

For training, we use cross-entropy loss and the Adam optimizer~\cite{Adam}. All models are trained for 150 epochs with a batch size of 256, an initial learning rate of $10^{-4}$, and weight decay of $10^{-4}$. Training uses randomly sampled 1\,s clips with no data augmentation. During inference, each utterance is segmented into non-overlapping 1\,s windows, and the output logits are averaged across windows, following~\cite{LEAF,AdaFE_Qiquan}. The checkpoint with the lowest validation loss is selected for testing. No data augmentation is applied to the raw audio at any stage. Each experiment is repeated three times, and we report the mean performance.
\begin{figure}[!ht]
  \centering
  \begin{subfigure}[b]{0.49\columnwidth}
    \centering
    \includegraphics[width=\linewidth]{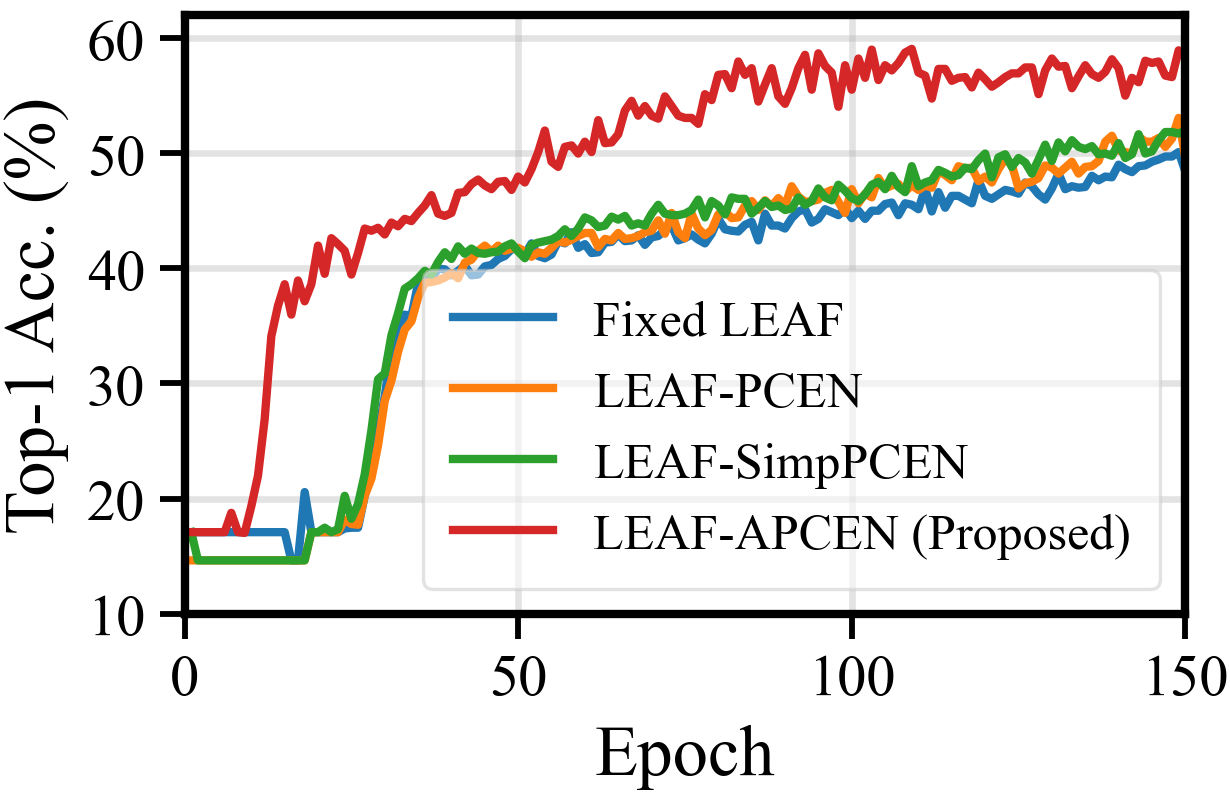}
    \subcaption{Speech Emotion Recognition in CREMA-D (clean).}
    \label{fig:result_1a}
  \end{subfigure}
  \hfill
  \begin{subfigure}[b]{0.49\columnwidth}
    \centering
    \includegraphics[width=\linewidth]{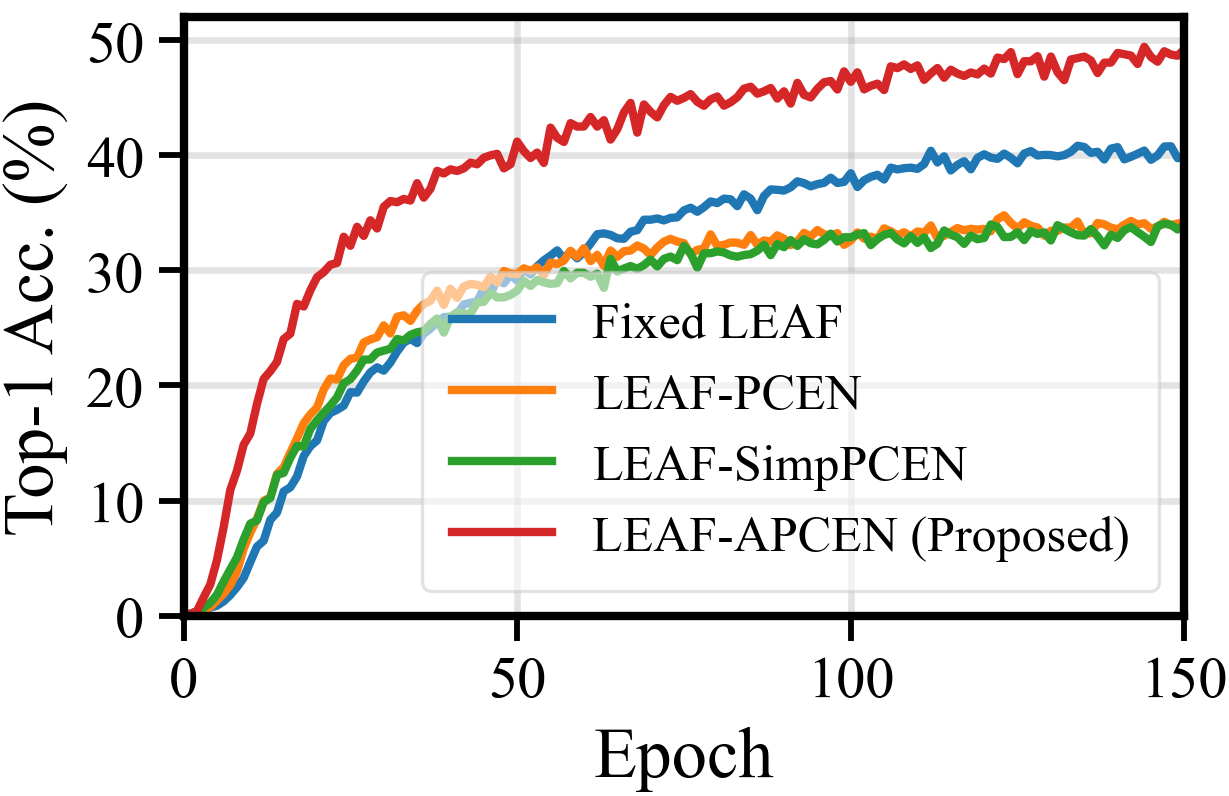}
    \subcaption{Speaker Identification in VoxCeleb1 (complex Env.).}
    \label{fig:result_1b}
  \end{subfigure}
  \vspace{-0.4em}
  \caption{Top-1 accuracy (\%) of the Fixed LEAF, LEAF-PCEN, LEAF-SimpPCEN, and LEAF-APCEN over various training epochs.}  
  \vspace{-3mm}
  \label{fig:result_1}
\end{figure}
\section{Evaluation Results and Discussion}
\label{sec:results}
\subsection{Results Analysis}
\label{sec:4.1}
Tables~\ref{tab:robust_res_clean} and~\ref{tab:robust_res_complex} summarize the test accuracy of the four front-end variants under clean and complex acoustic conditions.

From Table~\ref{tab:robust_res_clean}, LEAF-APCEN achieves the highest accuracy across all tasks except music genre classification. The improvements in speech-related tasks are substantial, reaching up to $8.40\%$ in emotion recognition and $14.47\%$ in speaker identification over the weakest baseline. This indicates that fixed or globally learned PCEN parameters are insufficient even in clean conditions, and dynamic adaptation across time and frequency is necessary.

Under complex acoustic conditions (Table~\ref{tab:robust_res_complex}), LEAF-APCEN consistently outperforms both fixed and learnable variants. Although the other models are trained on the same noisy and loudness-altered data, they lack the ability to adapt effectively to temporal energy variations during inference. In contrast, APCEN leverages temporal dependencies in the signal rather than fitting patterns from the training set, enabling superior robustness. For instance, in music genre classification, the fixed and learnable models suffer significant degradation under complex conditions, while LEAF-APCEN maintains stable performance and prevents the back-end classifier from collapsing.

We further tracked test accuracy across training epochs for different tasks and acoustic conditions, as shown in Fig.~\ref{fig:result_1}. The adaptive controller not only accelerates convergence from the beginning, but also achieves consistently higher final accuracy across datasets under both clean and complex conditions. These results demonstrate that the adaptive mechanism allows the model to better exploit input statistics and converge to more optimal operating points of the model, thereby improving both learning efficiency and generalization.

\begin{figure}[!ht]
  \centering
  \captionsetup[subfigure]{justification=centering,singlelinecheck=false,skip=2pt}
  \begin{subfigure}[t]{0.49\columnwidth}
    \centering
    \subcaption{\textbf{Before PCEN}}
    \includegraphics[width=\linewidth]{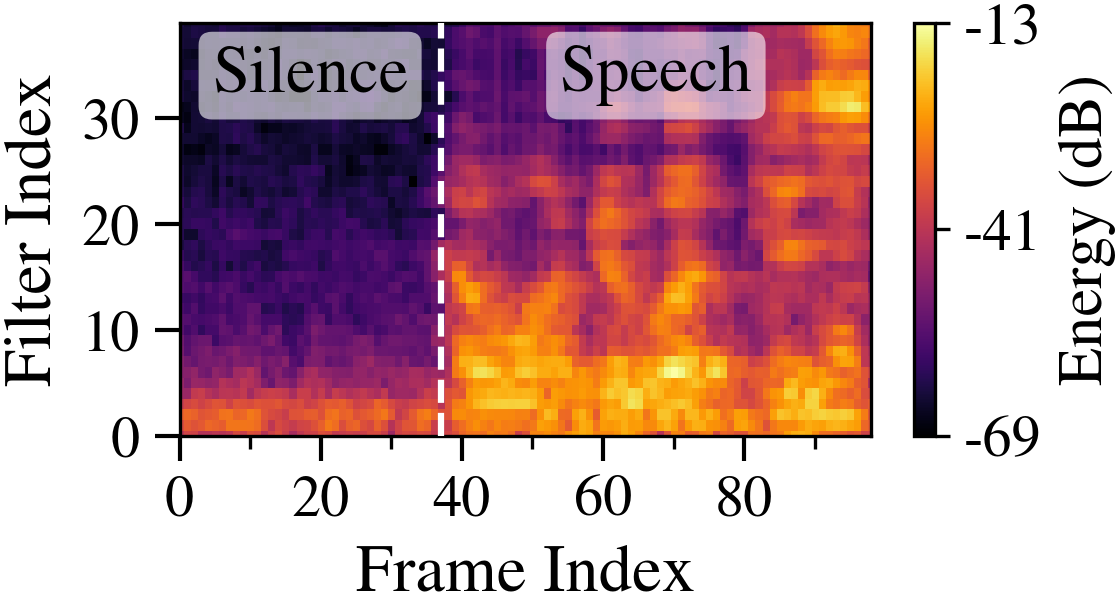}
    \label{fig:represent_before}
  \end{subfigure}\hfill
  \begin{subfigure}[t]{0.49\columnwidth}
    \centering
    \subcaption{\textbf{Fixed LEAF}}
    \includegraphics[width=\linewidth]{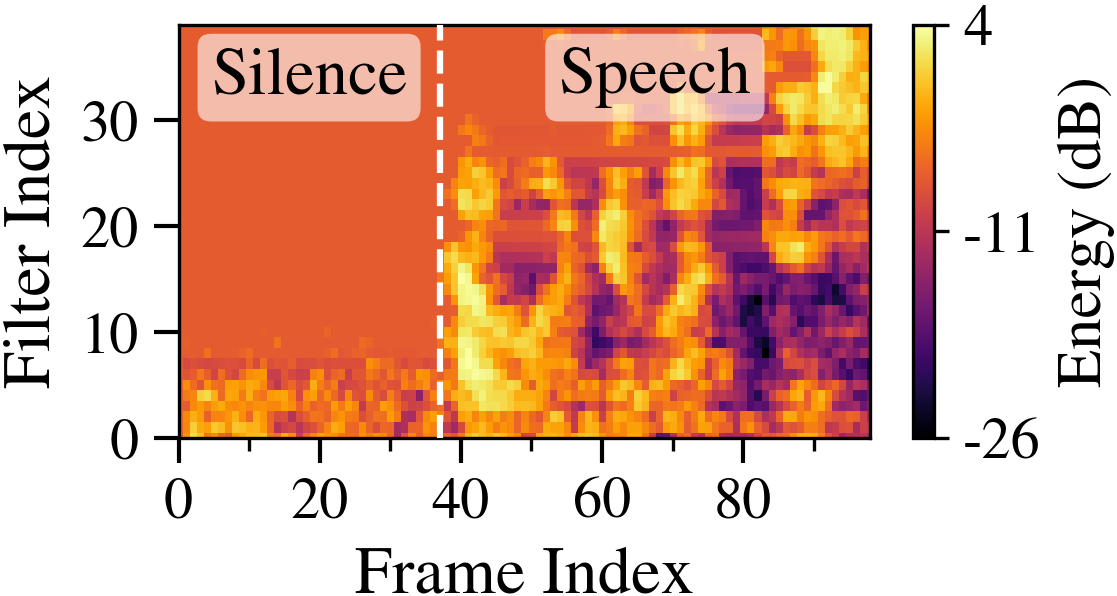}
    \label{fig:represent_fixed}
  \end{subfigure}
  \begin{subfigure}[t]{0.49\columnwidth}
    \centering
    \vspace{-2mm}
    \subcaption{\textbf{LEAF-PCEN}}
    \includegraphics[width=\linewidth]{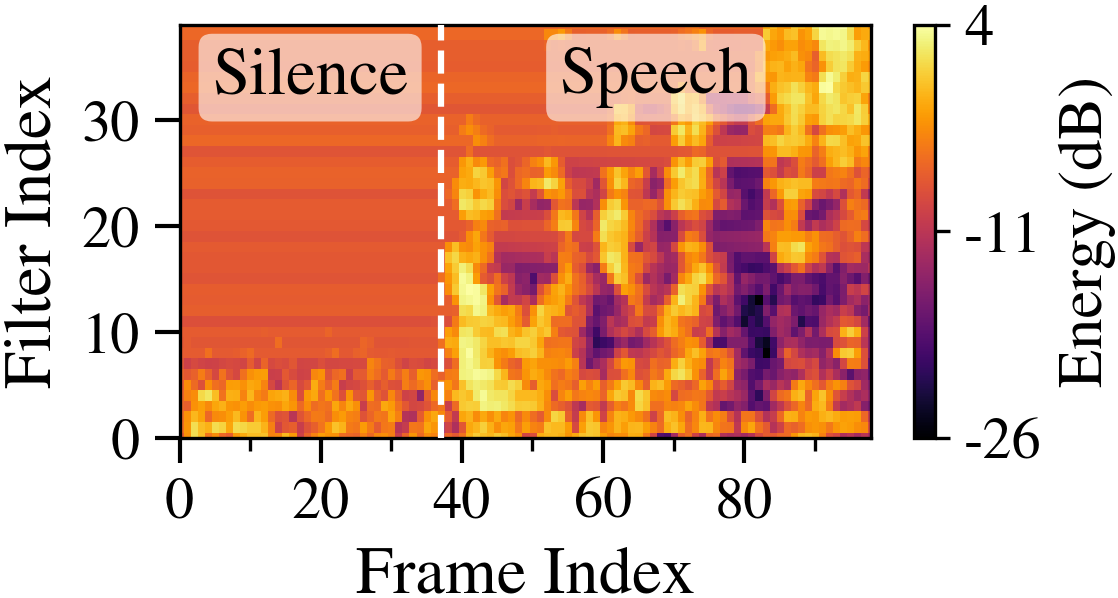}
    \label{fig:represent_LEAF}
  \end{subfigure}\hfill
  \begin{subfigure}[t]{0.49\columnwidth}
    \centering
    \vspace{-2mm}
    \subcaption{\textbf{LEAF-SimpPCEN}}
    \includegraphics[width=\linewidth]{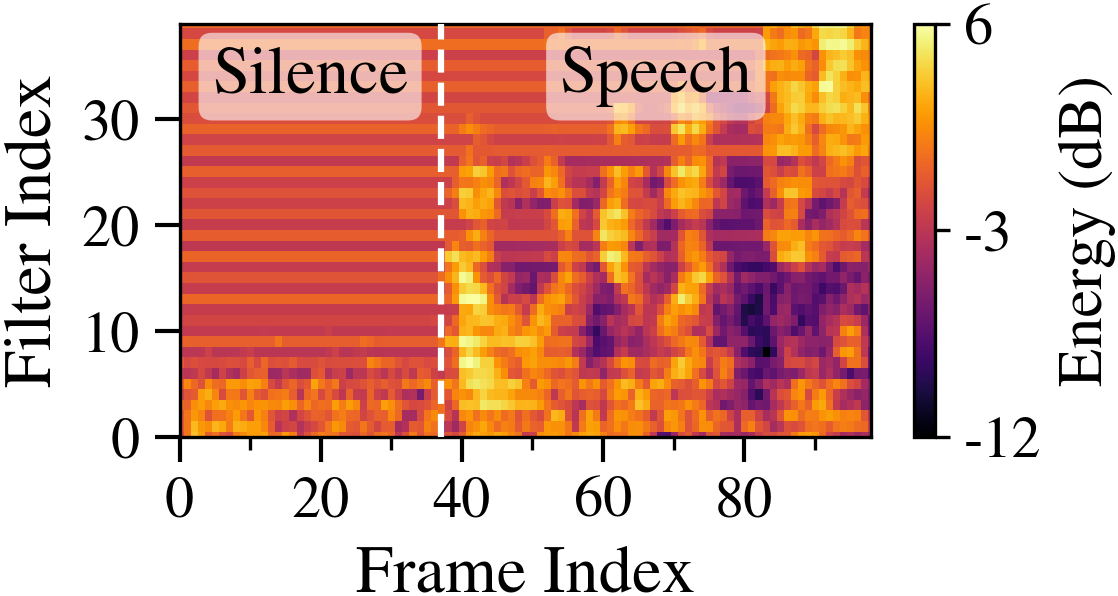}
    \label{fig:represent_SimpLEAF}
  \end{subfigure}
  \vspace{-2mm}
  \begin{subfigure}[t]{0.49\columnwidth}
    \centering
    \vspace{-2mm}
    \subcaption{\textbf{LEAF-APCEN}}
    \includegraphics[width=\linewidth]{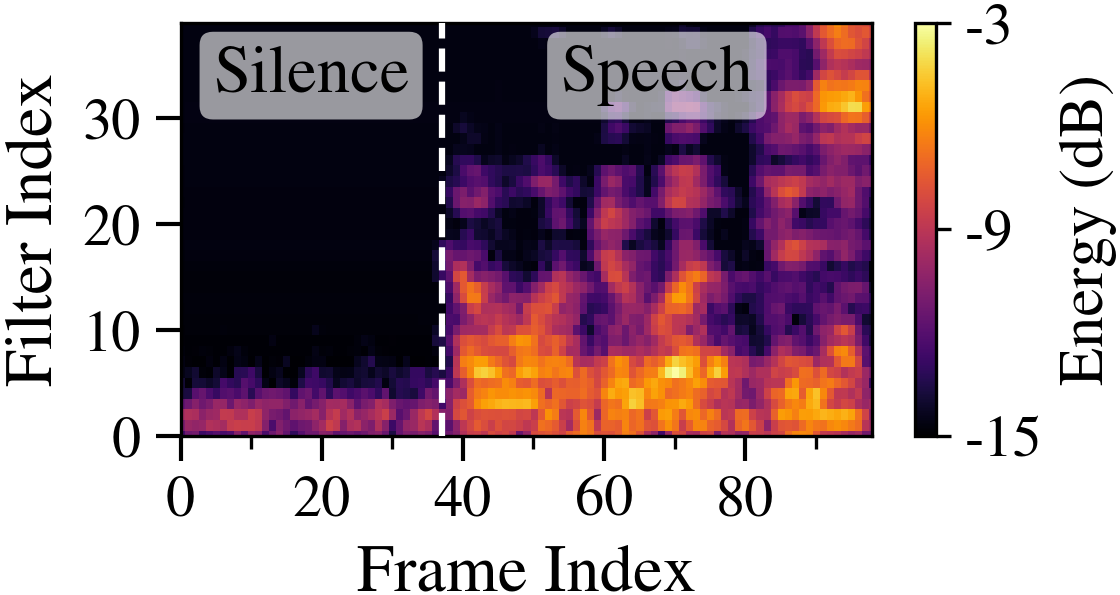}
    \label{fig:represent_ALEAF}
  \end{subfigure}\hfill
  \begin{subfigure}[t]{0.49\columnwidth}
    \centering
    \vspace{-2mm}
    \subcaption{\textbf{Adaptive PCEN Gain}}
    \includegraphics[width=\linewidth]{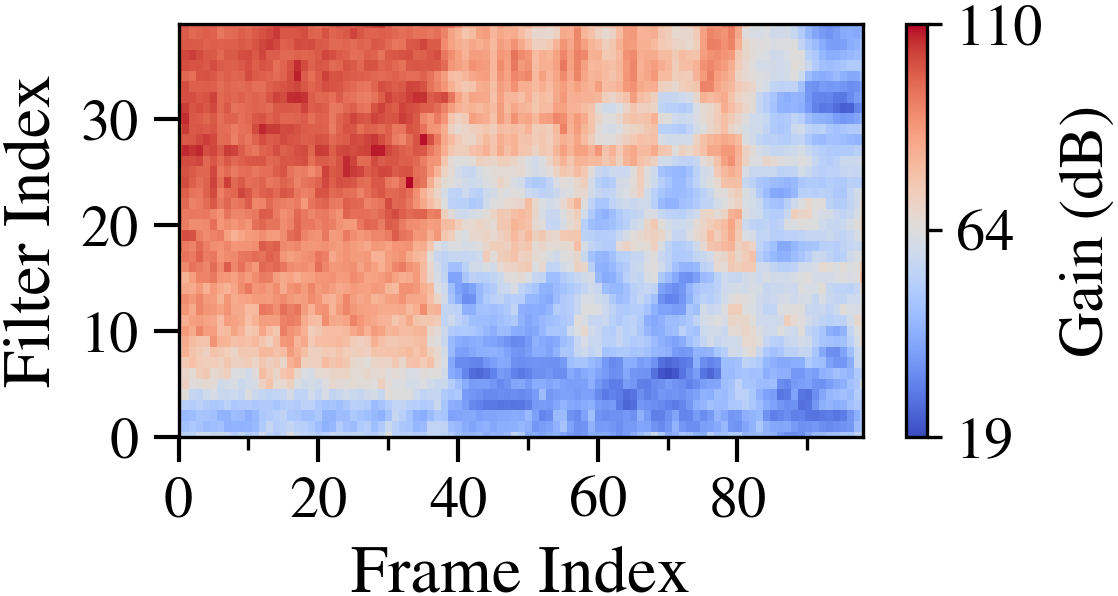}
    \label{fig:represent_A_gain}
  \end{subfigure}
  \vspace{-3mm}
  \caption{Representations of a CREMA-D test utterance with babble noise (SNR $\approx$1.5\,dB) across different stage and front-ends: (a) before PCEN (at the point indicated in Fig.~\ref{fig:overview}); (b–e) front-end representations obtained with different models; and (f) the adaptive time-varying gain from APCEN, computed as the ratio between input and output.}
  \label{fig:six_representation_3x2}
\end{figure}

\subsection{Adaptive Mechanism Analysis}
\label{sec:4.2}
To gain deeper insights into the performance benefits of LEAF-APCEN, we visualize and analyze representations produced by different front-ends. An utterance from the CREMA-D test set with babble noise (SNR $\approx$ 1.5 dB) is used, and its representations before and after each model’s PCEN are shown in Fig.~\ref{fig:represent_before}–\ref{fig:represent_ALEAF}.

Baseline models (Fixed LEAF, LEAF-PCEN, and LEAF-SimpPCEN) show that PCEN consistently rescales the energy while compressing its dynamic range. As illustrated in Fig.~\ref{fig:represent_fixed}–\ref{fig:represent_SimpLEAF}, Fixed LEAF and LEAF-PCEN yield similar outputs, reducing the original energy variation ($\approx$ 56 dB before PCEN) to about 30 dB, while LEAF-SimpPCEN compresses it further to $\approx$18 dB. However, these static or learnable variants fail to produce strong speech and silence contrast, limiting their ability to attend salient components in noisy conditions.

In contrast, LEAF-APCEN generates more discriminative features (Fig.~\ref{fig:represent_ALEAF}). In addition to dynamic range compression, it simultaneously enhances speech–silence contrast by amplifying informative regions and suppressing background interference. Furthermore, Fig.~\ref{fig:represent_A_gain} illustrates that the adaptive gain boosts low-energy regions and suppresses high-energy ones, preserving dynamic range compression while varying over time and frequency. Therefore, the adaptive mechanism in APCEN balances noise suppression and speech preservation, explaining LEAF-APCEN’s robustness in complex conditions with strong temporal and spectral variation.

\section{Conclusion}
\label{sec:conclude}
In this paper, we introduced LEAF-APCEN, a novel adaptive audio front-end framework designed for robust audio representation. Unlike conventional fixed or learnable front-ends with static parameterization, our approach incorporates a lightweight neural controller that dynamically tunes the two parameters of a SimpPCEN module during inference, enabling input-dependent adaptation. Experimental results demonstrate that the adaptive front-end significantly outperforms fixed and learnable front-ends across diverse audio classification tasks and substantially improves robustness under complex acoustic conditions. Further analysis shows that the controller generates time-varying parameters for energy normalization, thereby sharpening informative patterns while suppressing nuisance components. Overall, this work highlights adaptability as a promising direction for next-generation audio front-ends. Future work will extend the current single-channel design to multi-channel audio inputs, incorporating spatial information into downstream tasks and enabling joint temporal–spectral–spatial adaptation.
\vfill
\pagebreak
\bibliographystyle{IEEEbib}
\bibliography{refs}
\end{document}